\documentclass[aps,prl,10pt,twocolumn,showpacs,superscriptaddress]{revtex4-1}
\usepackage[latin1]{inputenc}
\usepackage[T1]{fontenc}
\usepackage[dvips]{epsfig}
\usepackage{amsmath}
\usepackage{amssymb}
\usepackage{braket}
\usepackage{bbm}

\usepackage{ulem}
\usepackage{color}
\definecolor{gold}{rgb}{1,0.75,0}

\begin{document}

\title{Particle self-bunching in the Schwinger effect in
  spacetime-dependent electric fields}
\author{F.~Hebenstreit}
\author{R.~Alkofer}
\affiliation{Institut f\"ur Physik, Karl-Franzens Universit\"at Graz, A-8010 Graz, Austria}
\author{H.~Gies}
\affiliation{Theoretisch-Physikalisches Institut, Friedrich-Schiller Universit\"at Jena \& Helmholtz-Institut Jena, D-07743 Jena, Germany}
\date{\today}

\begin{abstract}
  Non-perturbative electron-positron pair creation (Schwinger effect)  is
  studied based on the  {DHW} formalism {in 1+1
    dimensions}. An ab 
  initio calculation of the Schwinger effect in the presence of a simple
  space- and time-dependent electric field {pulse} is performed for the first
  time, allowing for the calculation of the time evolution of observable
  quantities such as the charge density, the particle number density or the
  total number of created particles. We {predict}  a new self-bunching effect of
  charges in phase space due to the spatial and temporal structure of the
  pulse.
\end{abstract}

\maketitle

\textbf{Introduction:} The vacuum of quantum electrodynamics (QED) is unstable
against the formation of many-body states in the presence of an external
electric field, manifesting itself as the creation of electron-positron pairs
\cite{Sauter:1931zz,Heisenberg:1935qt,Schwinger:1951nm}. This effect has been
a long-standing but still unobserved prediction as the generation of
{near-critical} 
field strengths $E_{\text{cr}}\sim10^{18}\,\mathrm{V/m}$ has not been
feasible so far. Due to the advent of a new generation of high-intensity laser
systems such as the European XFEL or the Extreme Light Infrastructure (ELI),
this effect might eventually become observable within the next decades.

Previous investigations of the Schwinger effect in the presence of
time-dependent electric fields
\cite{Brezin:1970xf,Popov:1972,Narozhnyi:1970uv,Alkofer:2001ik,Roberts:2002py,Blaschke:2005hs,
  Schutzhold:2008pz,Hebenstreit:2009km,Bulanov:2010ei,Dumlu:2010ua,Orthaber:2011cm},
space-dependent electric fields
\cite{Nikishov:1970br,Gies:2005bz,Dunne:2006ur,Kim:2007pm,Kleinert:2008sj} as
well as collinear electric and magnetic fields
\cite{Kim:2003qp,Tanji:2008ku,Ruffini:2009hg} led to a good understanding of
the general mechanisms behind the pair creation process by now. {However,} realistic
fields {of}
upcoming high-intensity laser
experiments showing both spatial and temporal variations have not been fully
considered yet. Only recently it became possible to study the Schwinger effect
in such realistic electric fields owing to recent theoretical progress as well
as due to the rapid development of computer technology. 
{Specifically,}
the {Dirac-Heisenberg-Wigner} phase-space formulation of QED in the presence of an external electric
field \cite{Vasak:1987um,BialynickiBirula:1991tx, Zhuang:1995pd,Ochs:1998qj}
(DHW formalism) has attracted interest again \cite{Hebenstreit:2010vz,Hebenstreit:2010cc,
  BialynickiBirula:2011uh}.
{It provides a real-time non-equilibrium formulation of the quantum
  production process. Also,}
a one-to-one
mapping between the {DHW} function (phase-space formalism) and the
one-particle distribution function (quantum kinetic formalism) {exists}
in the limit of a spatially homogeneous, time-dependent electric field.

The Schwinger effect in the presence of an arbitrary {spacetime-dependent}
electric field {is} properly described by the 
{DHW} formalism {in the form of} a partial differential equation (PDE) system
for the irreducible components of the {DHW} function. The numerical solution
of the PDE system allows for the calculation of any observable quantity in
terms of the irreducible components.

In the present work, we consider a simple model for a sub-attosecond high-intensity laser pulse in standing wave mode with finite extension. In the focus of the beam, pair production along the direction of the electric field gives the dominant contribution to the Schwinger effect. Ignoring particle momenta orthogonal to this dominant direction, the system reduces to a 1+1 dimensional setting, which is studied for the first time here and solved numerically \cite{Hebenstreit:2011}.

\textbf{Formalism:} 
{Following the fundamental work of \cite{BialynickiBirula:1991tx}},
we start with the gauge-invariant equal-time commutator of two Dirac field operators:
\begin{equation}
  \Phi(x,y,t):=\mathcal{U}(x,y)[\bar{\Psi}(x-y/2,t),\Psi(x+y/2,t)] \ ,
\end{equation}
with $x$ denoting the center-of-mass and $y$ the relative coordinate. Here,
the Wilson-line factor which ensures gauge invariance is chosen along {a} straight line:
\begin{equation}
  \mathcal{U}(x,y)=\exp\left(-ie\int_{-1/2}^{1/2}\mathrm{d}\xi A(x+\xi y,t)\,y\right) \ .
\end{equation}
The vector potential $A(x,t)$ is treated as classical mean field, i.e. photon fluctuations are neglected. This approximation is well justified for the pair-production process in QED.
Tree-level radiation reactions which might play a sizable role for strong fields according to recent investigations \cite{Bell:2008zzb,Fedotov:2010ja,Bulanov:2010gb} are also neglected in this work.

Taking the vacuum expectation value $\bra{\Omega}\Phi(x,y,t)\ket{\Omega}$,
{we trade $y$} for a kinetic momentum variable $p$ {by} 
a Fourier transformation. 
This defines the {DHW}
function:
\begin{equation}
  \mathcal{W}(x,p,t) 
  :=\frac{1}{2}\int{\mathrm{d}y\,e^{-ipy}\bra{\Omega}\Phi(x,y,t)\ket{\Omega}} \ .
\end{equation}
Due to the fact that $\mathcal{W}(x,p,t)$ 
{is in the} Dirac {algebra}, it may be decomposed
in terms of its Dirac bilinears:
\begin{equation}
  \mathcal{W}(x,p,t)=\frac{1}{2}[\mathbbm{s}+i\gamma^5\mathbbm{p}+\gamma^\mu\mathbbm{v}_\mu] \ ,
\end{equation}
with irreducible components transforming as scalar
$\mathbbm{s}(x,p,t)$, pseudoscalar $\mathbbm{p}(x,p,t)$ and vector
$\mathbbm{v}_\mu(x,p,t)$. {For brevity, these
  components will later on collectively be denoted as $\mathbbm{w}(x,p,t)$.} The
derivation of the corresponding equations of motion {follows that }
in 3+1 dimensions
\cite{BialynickiBirula:1991tx,Hebenstreit:2011} and yields the following
hyperbolic PDE system:
\begin{alignat}{8}
  \label{DHW1}
  &[\tfrac{\partial}{\partial t}+\Delta]\,\mathbbm{s}\ &\  &\  &\ -&\ 2p\,\mathbbm{p}\ &\ =\ &\ &\ & 0 &\ ,  \\ 
  &[\tfrac{\partial}{\partial t}+\Delta]\,\mathbbm{v}_0\ &\ +&\ \tfrac{\partial}{\partial x}\mathbbm{v}\ &\  &\  &\ =\ &\ &\ & 0&\  , \\ 
  &[\tfrac{\partial}{\partial t}+\Delta]\,\mathbbm{v}\ &\ + &\ \tfrac{\partial}{\partial x}\mathbbm{v}_0\ &\ &\  &\ =\ &\ -&2m & \mathbbm{p}&\  , \\
  &[\tfrac{\partial}{\partial t}+\Delta]\,\mathbbm{p}\ &\ &\ & + &\ 2p\,\mathbbm{s}&\ =\ &\ &2m & \mathbbm{v}&\ ,
  \label{DHW2}
\end{alignat}
with the pseudo-differential operator
\begin{equation}
  \Delta(x,p,t) = \ e\int_{-1/2}^{1/2}{\mathrm{d}\xi\,
  E\big(x+i\xi \tfrac{\partial}{\partial p},t\big)\tfrac{\partial}{\partial p}} \ .
\end{equation}
Along with $\omega(p)=\sqrt{m^2+p^2}$, the appropriate vacuum initial
  conditions at asymptotic times $t_\mathrm{vac}\to-\infty$ are
\begin{equation}
  \mathbbm{s}_\mathrm{vac}(p)=-\frac{m}{\omega(p)} \quad \mathrm{and} \quad 
  \mathbbm{v}_\mathrm{vac}(p)=-\frac{p}{\omega(p)} \ .
\end{equation}
The irreducible components are not {directly} observable, however, they
constitute the observable quantities which can be derived from Noether's
theorem. For our purpose, the charge $\mathcal{Q}(t)$ as well as the energy of
the Dirac particles $\mathcal{E}(t)$ are of special interest:
\begin{alignat}{4}
  \mathcal{Q}(t)&\,=\,&e&\int{\mathrm{d}\Gamma\,\mathbbm{v}_0(x,p,t)} \ , \\ 
  \mathcal{E}(t)&\,=\,&&\int{\mathrm{d}\Gamma\,[m\,\mathbbm{s}(x,p,t)+p\,\mathbbm{v}(x,p,t)]}\ ,
\end{alignat}
with $\mathrm{d}\Gamma=\mathrm{d}x\mathrm{d}p/(2\pi)$ denoting the phase space
volume element. The integrands $q(x,p,t)=\mathbbm{v}_0(x,p,t)$ and
  $\epsilon(x,p,t)=[m\,\mathbbm{s}(x,p,t)+p\,\mathbbm{v}(x,p,t)]$ can be
  viewed as pseudo-charge density and pseudo-energy density, respectively.
Due to the fact that we are considering a quantum theory, it is more
appropriate to consider the momentum space marginal distributions:
\begin{eqnarray}
  q(p,t)&:=&\int{\frac{\mathrm{d}x}{(2\pi)}\,q(x,p,t)} \ , \\
  \epsilon(p,t)&:=&\int{\frac{\mathrm{d}x}{(2\pi)} 
  \,\big[m\,\mathbbm{s}(x,p,t)+p\,\mathbbm{v}(x,p,t)\big]} \ .
\end{eqnarray}
{Requiring} that the total energy of the Dirac particles should be calculable
by integrating a particle number pseudo-distribution $n(x,p,t)$ times the
one-particle energy $\omega(p)$, it is also {useful} to introduce the momentum
space particle number densities
\begin{eqnarray}
  n(p,t)&:=&\int{\frac{\mathrm{d}x}{(2\pi)}\,n(x,p,t)} \ ,
\end{eqnarray}
with
\begin{equation}
  n(x,p,t)=\frac{m[\mathbbm{s}(x,p,t)\!-\mathbbm{s}_\mathrm{vac}(p)]+ 
  p\,[\mathbbm{v}(x,p,t)\!-\mathbbm{v}_\mathrm{vac}(p)]}{\omega(p)}.
\end{equation}
{The vacuum subtractions account for a normalization of the density
  relative to the vacuum Dirac sea.}
Accordingly, the total number of created particles reads:
\begin{equation}
  N(t)=\int{\mathrm{d}p\,n(p,t)} \ .
\end{equation}

The PDE system Eqs.~(\ref{DHW1}) -- (\ref{DHW2}) calls for further rewritings or even approximations as arbitrarily high momentum derivatives have to be taken into account in general:

\textit{(a) Full solution in conjugate space:} As the momentum $p$
appears linearly in the PDE system Eqs.~(\ref{DHW1}) -- (\ref{DHW2}), we can
transform these equations to conjugate $y$ space. 
 As a consequence,  $\Delta(x,p,t)$ transforms into a function of $y$ 
as well:
\begin{equation}
  \int{\frac{\mathrm{d}p}{(2\pi)}e^{ipy}\Delta(x,p,t)}=-iey\int_{-1/2}^{1/2}{\mathrm{d}\xi E(x+\xi y,t)} \ ,
\end{equation}
resulting in an {\bf{exact}}, first order PDE system. 

\textit{(b) Leading  order derivative expansion:} The simplest approximation is to expand $\Delta(x,p,t)$ in a series with respect to the spatial variable. Requiring that \cite{Hebenstreit:2010vz}:
\begin{equation}
  \big|E(x,t)\tfrac{\partial  \mathbbm{w}(x,p,t)]}{\partial p}\big|\gg  \tfrac{1}{24}\big|E''(x,t)\tfrac{\partial^3  \mathbbm{w}(x,p,t)]}{\partial p^3}\big| \ ,
\end{equation}
it is well justified to neglect the higher derivatives:
\begin{equation}
  \Delta(x,p,t)\simeq eE(x,t)\tfrac{\partial}{\partial p} \ ,
\end{equation}
yielding an {\bf{approximate}}, first order PDE system.

\textit{(c) Local density approximation:} Approximations can also be constructed on the level of the marginal distribution $n(p,t)$. Given an
electric field $E(x,t)=E_0g(x)h(t)$, and assuming that the spatial variation scale is much larger than the Compton wavelength $\lambda\gg\lambda_C$, it is well justified to locally describe the Schwinger effect at any point $x$ independently. We then define the particle number quasi-distribution in local density approximation as:
\begin{equation}
  n_\mathrm{loc}(x,p,t):=2\mathcal{F}(p,t;x) \ .
\end{equation}
$\mathcal{F}(p,t;x)$ denotes the one-particle distribution function which is found by solving the quantum Vlasov equation \cite{Schmidt:1998vi,Bloch:1999eu} at any {fixed} point $x_{\text{fixed}}$ for a time-dependent electric field $E(t)=E_0g(x_{\text{fixed}})h(t)$. Accordingly: 
\begin{eqnarray}
  n_\mathrm{loc}(p,t)&:=&\int{\frac{\mathrm{d}x}{(2\pi)}\,n_\mathrm{loc}(x,p,t)} \ .
\end{eqnarray}

\begin{figure}[b!]
  \centering
  \includegraphics[width=0.44\textwidth]{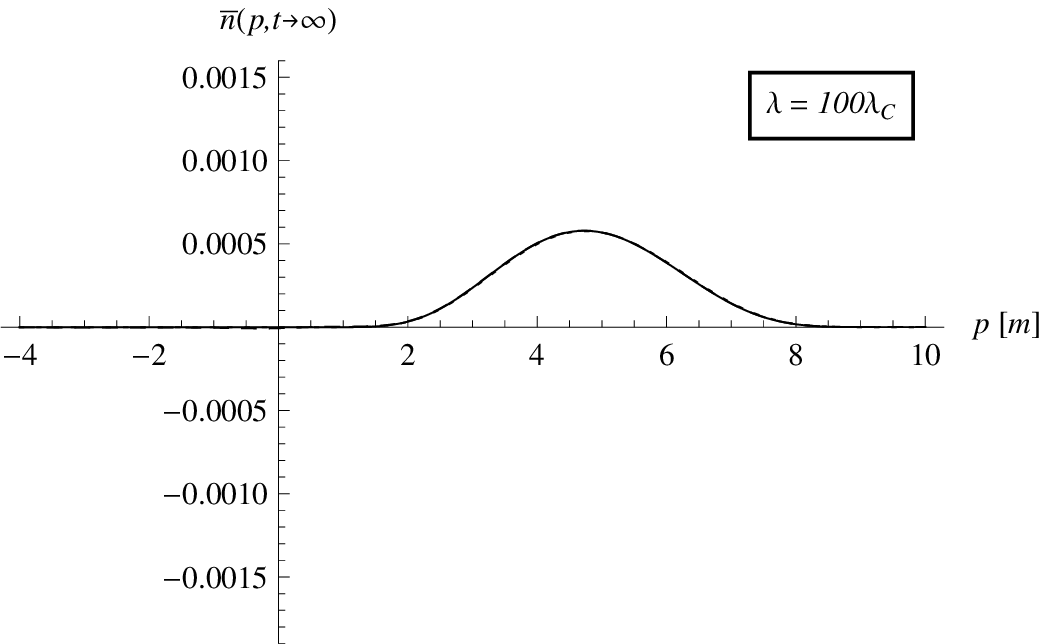}
  \includegraphics[width=0.44\textwidth]{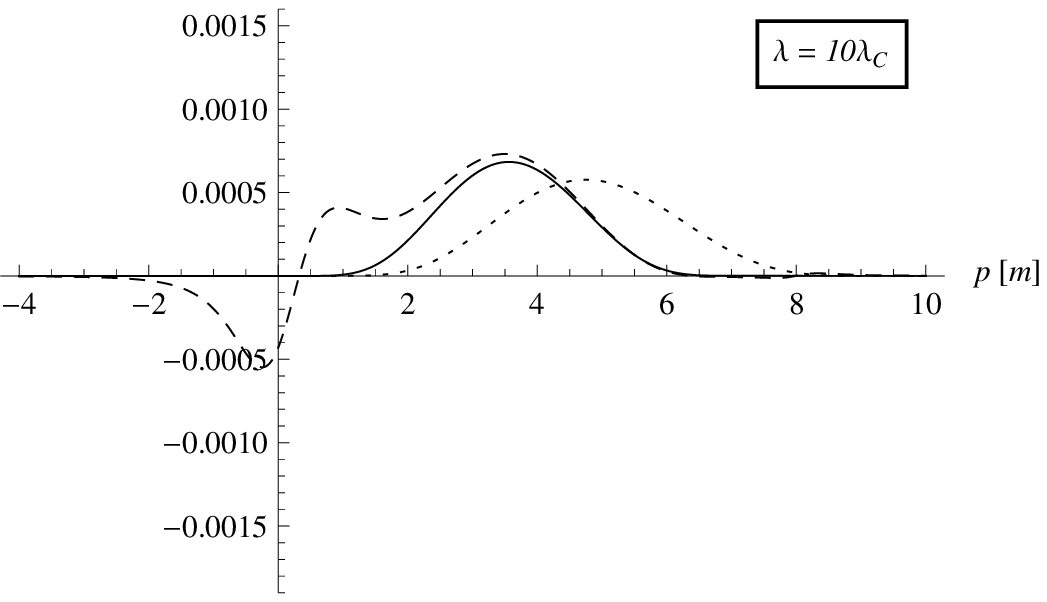}
  \includegraphics[width=0.44\textwidth]{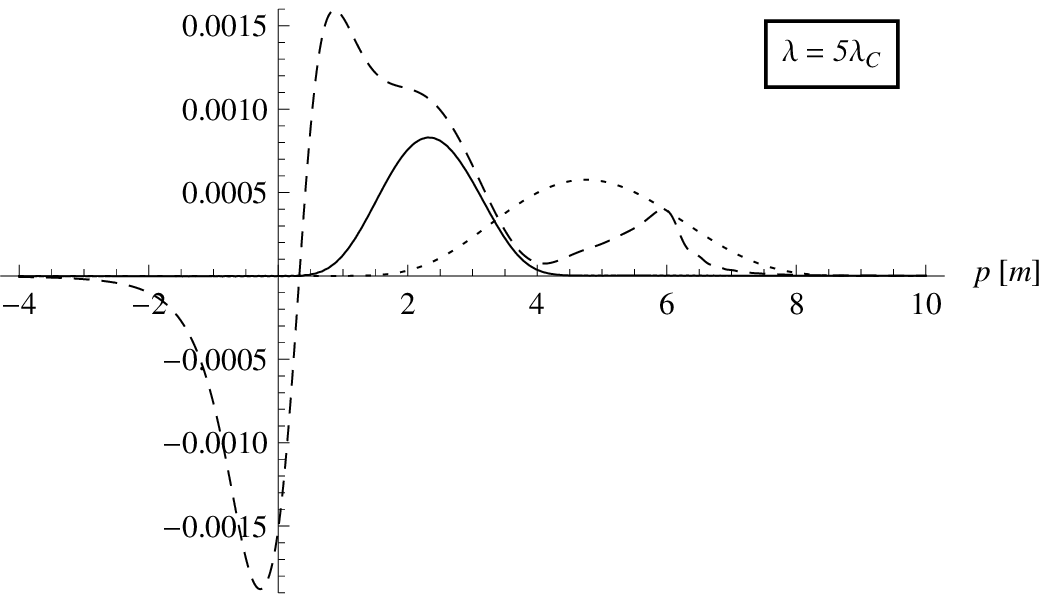}
  \caption{Comparison of $\bar{n}(p,t\to\infty)$ for the full solution (solid) with the l.o.~derivative expansion (dashed) and the local density approximation (dotted) for $\tau=10/m$, {$E_0=0.5E_{\text{cr}}$}.}
  \label{fig_approx}
\end{figure}

\textbf{Results:} Our idealized model for a spatially and temporally well-localized laser pulse in a standing
wave mode is parameterized by the electric field:
\begin{equation}
  \label{elfield}
  E(x,t)=E_0\exp\big(-\tfrac{x^2}{2\lambda^2}\big)\operatorname{sech}^2\big(\tfrac{t}{\tau}\big) \ ,
\end{equation}
with $\tau$ and $\lambda$ denoting the characteristic time and length scale,
respectively. We {choose} the parameters $\tau = 10/m$,
$E_0=0.5E_{\text{cr}}$ in this investigation{, corresponding to an
  intense sub-attosecond pulse}. As the spatial
  extent as well as the total energy of the electric field of the pulse
  decrease with $\lambda$, if all other parameters are held fixed, it is
convenient to disentangle this trivial scaling effect and investigate scaled
quantities for better comparability:
\begin{equation}
  \bar{n}(p,t):=\frac{n(p,t)}{\lambda} \qquad \mathrm{and} \qquad \bar{N}(t):=\frac{N(t)}{\lambda} \ .
\end{equation}

\textit{Full solution vs. approximations:} In Fig.~\ref{fig_approx} we compare
the asymptotic value $\bar{n}(p,t\to\infty)$ {of} the full solution with the leading order derivative expansion as well as with the local density approximation for different values of $\lambda$. The difference between the various results is rather small for broader pulses. As the various approximations are in good agreement with the full solution, the pair creation process can indeed be considered as taking place at any point $x$ independently in this regime. For decreasing $\lambda$, however, the various results differ substantially.

As expected, the leading order derivative expansion becomes worse for small
$\lambda$. Whereas a previous study of higher derivative terms signalled a
potential failure at large momenta \cite{Hebenstreit:2010vz}, we here observe
a breakdown of this approximation for small momenta $p/m\to0$. For larger
$\lambda$, the dominant momenta are still well approximated, but for $\lambda$
approaching $\lambda_{\text{C}}$, the truncation artefacts overwhelm the
physical values. Also the fact that the particle density
$\bar{n}(p,t\to\infty)$ acquires negative values in the derivate
expansion signals a clear breakdown of this approximation for small
momenta. The local density approximation fails in a different respect: The
peak momentum of the full solution is shifted to smaller values for decreasing
$\lambda$ {which is not reflected by} 
the local density approximation. 

\begin{figure}[b!]
  \centering
  \includegraphics[width=0.44\textwidth]{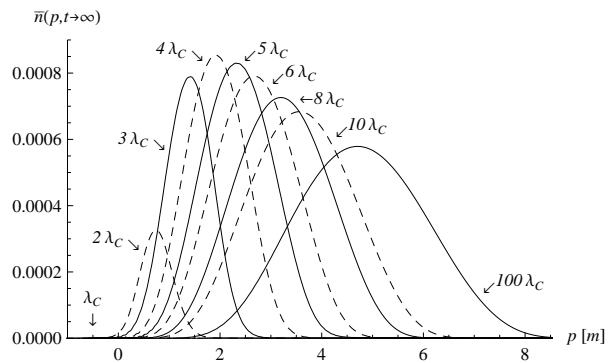}
  \caption{Comparison of $\bar{n}(p,t\to\infty)$ for
    $\tau=10/m${, $E_0=0.5E_{\text{cr}}$} and different values
    of $\lambda$. Note that $\bar{n}(p,t\to\infty)=0$ for
    $\lambda=\lambda_C$.}
  \label{fig_lambda}
\end{figure}

\textit{Particle number density:} In Fig.~\ref{fig_lambda}, we investigate the behavior of the full solution $\bar{n}(p,t\to\infty)$ for different values of $\lambda$. A decreasing $\lambda$ involves a shift of the peak momentum to a smaller value: The value of the acceleration by the electric field depends on the actual position such that the field excitations feel a varying acceleration when moving through the electric field. Accordingly, the field excitations are less accelerated for narrow pulses.

Morover, the shape of $\bar{n}(p,t\to\infty)$ becomes higher and narrower for decreasing $\lambda$, at least for $\lambda\gtrsim4\lambda_C$. This is  a self-bunching effect caused by the spatial inhomogeneity: Excitations which are created with high momenta are accelerated for a shorter period as they leave the field rapidly. By contrast, excitations which are created with small momenta stay longer inside the field and are accelerated for a longer period. Accordingly, the created particles are bunched into a smaller phase space volume.

\begin{figure}[t!!]
  \centering
  \includegraphics[width=0.45\textwidth]{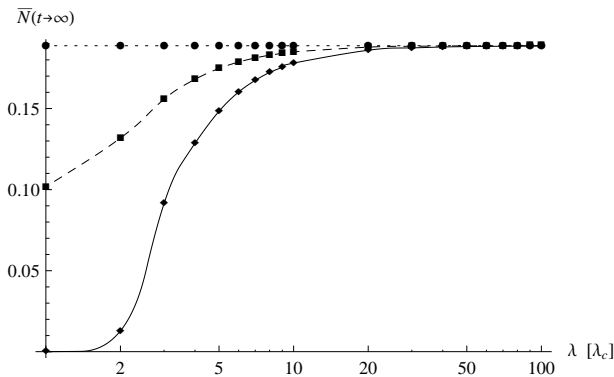}
  \caption{Comparison of $\bar{N}(t\to\infty)$ for the full solution (solid)
    with the l.o.~derivative expansion (dashed) and the local density
    approximation (dotted) for $\tau=10/m$, {$E_0=0.5E_{\text{cr}}$}.}
  \label{fig_number}
\end{figure}

For $\lambda\lesssim4\lambda_C$, however, the height of
$\bar{n}(p,t\to\infty)$ decreases again as more and more field excitations
gain too {little} energy in order to finally turn into real particles.  For
$\lambda=\lambda_C$, the energy content of the electric field is ultimately so
small that none of the vacuum fluctuations eventually turns into real
particles. This observation is in good agreement with previous studies of
space-dependent electric fields $E(x)$ \cite{Nikishov:1970br,Gies:2005bz,Dunne:2006ur}: The pair creation process is
expected to terminate once the work done by the electric field over its
spatial extent is smaller than twice the electron mass. {As}
the pair creation process 
occurs at time scales of the order of the
Compton time $t_C=1/m$, which is smaller than the 
time
scale of the electric pulse $\tau=10/m$, this estimate should be
reasonable in our case as well. The corresponding estimate for the pair
  creation process to terminate for $E_0=0.5E_{cr}$ is in fact in good
agreement with our results:
\begin{equation}
  \label{terminate}
  \lambda<\frac{E_{cr}}{E_0}\sqrt{\frac{2}{\pi}}\lambda_C\simeq1.6\lambda_C \ .
\end{equation}
\textit{Number of created particles:} In Fig.~\ref{fig_number} we compare the asymptotic value $\bar{N}(t\to\infty)$ obtained from the full solution with the leading order derivative expansion as well as with the local density approximation for different values of $\lambda$. Again, we observe good agreement between the full solution and the various approximations for large $\lambda$, however, substantial deviation for small $\lambda$. Most notably, only the full solution shows the sharp drop of $\bar{N}(t\to\infty)$ for small $\lambda$ in accordance with Eq.~(\ref{terminate}).

\textbf{Conclusions:} We have presented 
an ab initio
{real-time} calculation of the Schwinger effect in the presence of a
simple space- and time-dependent electric field pulse in 1+1
dimensions, showing various remarkable features:

Most notably, we observe a new self-bunching effect in phase space which can
naturally be interpreted in terms of the space and time evolution of the
quantum excitations. The pair creation process eventually terminates 
{for spatially small pulses} once
the work done by the electric
field is too small in order to provide the rest mass of an electron-positron
pair. Whereas the derivative expansion is quantitatively able to signal these
self-bunching effects, the local density approximation fails to describe these
important properties.

These results suggest further studies of the Schwinger effect in
realistic space- and time-dependent electric fields
{in} 3+1 dimensions. The 
goal
 is to consistently describe the Schwinger effect beyond the mean
field level by taking into account photon corrections to the background
electric field and subsequent collision and radiation processes. {In the
  long run, we expect the self-bunching effect to play an important role in
  the generation of taylored electron/positron beams.}

{\bf{Acknowledgments:}} FH is supported by the DOC program of the \"OAW and by the FWF doctoral program DK-W1203. HG acknowledges support by the DFG through grants SFB/TR18, and Gi 328/5-1.

\end{document}